\documentclass[12pt,aps,prb,superscriptaddress,amsmath,amssymb]{revtex4-1}

\usepackage{graphicx,subfigure}
\usepackage{dcolumn}
\usepackage{bm}
\usepackage{epsf,color}
\usepackage{fancyhdr}
\usepackage{amsfonts}
\usepackage{float, rotating, rotfloat}
\usepackage[section]{placeins} 
\usepackage{gensymb} 
\usepackage{setspace}
\usepackage{amsmath}

\usepackage{algcompatible, algpseudocode}
\usepackage[floatrow]{trivfloat}
\trivfloat{algorithm}

\usepackage{etoolbox}
\makeatletter
\preto{\@verbatim}{\topsep=0pt \partopsep=0pt }
\makeatother

\input vmc.def
\input dmc.def
\input projectors.def

\def\erf{{\rm erf}}

\def\psii{\psi_{i}}
\def\psij{\psi_{j}}
\def\psiij{\psi_{ij}}
\def\psisq{\psi^2}

\def\psir1pN{\psi({\bf r}'_1,{\bf r}_2,...,{\bf r}_N)}

\def\Psir1pN{\Psi({\bf r}'_1,{\bf r}_2,...,{\bf r}_N)}

\def\PsiTr1pN{\PsiT({\bf r}'_1,{\bf r}_2,...,{\bf r}_N)}

\def\EL{E_{\rm L}}
\def\ELRvec{E_{\rm L}(\Rvec)}
\def\ELkRvec{E_{{\rm L},k}(\Rvec_k)}

\def\ET{E_{\rm T}}
\def\ETk{E_{{\rm T},k}}

\def\rvec{{\bf r}}

\def\Rvec{{\bf R}}

\def\X2bar{\overline{X^2}}

\def\dw{\Delta w\left(\Rvec,\taueff\right)}
\def\dwk{\Delta w_{\rm k} \left(\Rvec_k,\taueffk\right)}

\def\beq{\begin{eqnarray}}
\def\eeq{\end{eqnarray}}

\def\br{{\bf r}}
\def\lmax{{l_{\rm max}}}
\def\gtNL{\tilde{g}^{\rm NL}}
\def\ftNL{\tilde{f}^{\rm NL}}
\def\tNL{t^{\rm NL}}

\def\SR{S\left(\Rvec,\tau\right)}
\def\SRp{S\left(\Rvecp,\tau\right)}
\def\fRt{f\left(\Rvec,\tau\right)}
\def\fRteff{f\left(\Rvec,\taueff\right)}
\def\taueff{\tau_{\rm eff}}
\def\taueffk{\tau_{{\rm eff},k}}

\def\rveci{\rvec_{\rm i}}
\def\rvecip{\rvec'_{\rm i}}
\def\rvecj{\rvec_{\rm j}}

\def\rij{\lvert \rveci - \rvecj \rvert}

\def\Nfrag{M}

\def\sumk{\sum^\Nfrag_{k=1}}
\def\prodk{\prod^\Nfrag_{k=1}}

\def\Ppropf{P_{\rm prop}(\Rvecipp,\Rveci,\tau)}
\def\Ppropb{P_{\rm prop}(\Rveci,\Rvecipp,\tau)}

\begin{document}

\title{Reducing the time-step errors in diffusion Monte Carlo}

\author{Tyler A. Anderson\footnote{taa65@cornell.edu}}
\affiliation{
Laboratory of Atomic and Solid State Physics,\\
Cornell University, Ithaca, NY 14853, United States of America.}
\author{Manolo C. Per\footnote{Manolo.Per@data61.csiro.au}}
\affiliation{
CSIRO Data61, \\ 
Clayton, VIC 3168, Australia.}
\author{C. J. Umrigar\footnote{CyrusUmrigar@cornell.edu}}
\affiliation{
Laboratory of Atomic and Solid State Physics,\\
Cornell University, Ithaca, NY 14853, United States of America.}

\begin{abstract}
We modify the reweighting factor of the projector used in diffusion Monte Carlo to reduce the time-step error
of the total energy.  Further, we present a reweighting
scheme that has the desirable feature that it is exactly size-consistent, i.e, the energy of a system containing
widely separated fragments is the same as the sum of the energies of the individual fragments.
The practical utility of the latter improvement is that it reduces the time-step error
of the binding energies of some weakly interacting systems.
\end{abstract}

\maketitle

\def\br{{\bf r}}
\def\lmax{{l_{\rm max}}}
\def\gtNL{\tilde{g}^{\rm NL}}
\def\ftNL{\tilde{f}^{\rm NL}}
\def\tNL{t^{\rm NL}}
\def\fmix{f_{\rm mix}(\Rvec)}
\def\VR{{\bf V}(\Rvec)}
\def\nrand{\bf \mathcal{N}}
\def\nrandi{{\bf \mathcal{N}}_i}
\def\ith{i^{\rm th}}
\def\fUNRR{f_{\rm UNR}(\Rvec,\tau)}
\def\sumi{\sum^N_i}
\def\Pacc{P_{\rm acc}(\Rvecipp,\Rveci,\tau)}
\def\di{\left\lvert \rvecipp - \rveci -\vbarveciRi \tau \right\rvert}
\def\lv{\left\lvert}
\def\rv{\right\rvert}

\section{Introduction}
\label{sec:intro}

Quantum Monte Carlo (QMC) methods are among the most accurate electronic structure methods.
Real-space QMC methods~\cite{BecSor-Book-17,ReyCepAldLes-JCP-82}, i.e. QMC methods wherein the Monte Carlo
walk is performed in real space, have the unique advantage compared to all other electronic structure
methods that they work directly in a complete basis.
The diffusion Monte Carlo (DMC) method is by far the most commonly used real-space QMC method.~\footnote{In practice
DMC calculations employ the fixed-node approximation, which introduces some dependence on the basis.
However, this basis set dependence is much weaker than in conventional quantum chemistry methods.}

DMC employs the exponential projector, $e^{\tau(\ET-\hat{H})}$, where $\hat{H}$ is the
Hamiltonian and $\ET$ an estimate of the ground state energy, to project onto the ground state wave function.
DMC calculations can be very computationally expensive because the exact expression for the exponential projector in real space is unknown, necessitating the use
of approximate projectors which result in a time-step error which vanishes in the small time-step limit, $\tau \to 0$.
In practice one must use either a very small $\tau$ or several small $\tau$'s and extrapolate to the $\tau \to 0$ limit.
For fixed computer time, the statistical errors of expectation values of observables go as $1/\sqrt{\tau}$ so
long calculations are needed to achieve the desired precision when $\tau$ is small.

This paper has two interrelated goals.  
First a new reweighting factor, obtained by satisfying five requirements which we argue should be satisfied by the exact reweighting factor, is
presented that greatly reduces the time-step error of the total energy.
Second, we present another reweighting factor which, for a composite system consisting of multiple parts, reduces to the
product of the reweighting factors of each individual system in the limit that the systems are widely separated
and the trial wave function is a product of the individual trial wave functions of each system.
The practical utility of this choice is that the time-step error of the binding energy of weakly bound systems
is smaller than the time-step errors of the energies of both the compound system and the individual systems.

The rest of this paper is organized as follows. In Sec.~\ref{sec:DMC} we briefly review the DMC algorithm.
In Sec.~\ref{sec:rewt_factor} we present our reweighting factor which results in much smaller time-step errors
than those commonly used.  In Sec.~\ref{sec:fragment} we discuss modifications needed
to ensure that the size-consistency error is zero.
This helps make the time-step errors of binding energies of weakly bound systems smaller than the
time-step errors of the individual total energies, in most cases.
In Sec.~\ref{sec:results} we demonstrate the reduction in the time-step errors for several systems.
In Sec.~\ref{sec:conclusions} we present our conclusions.

\section{Diffusion Monte Carlo}
\label{sec:DMC}

We now briefly describe the current standard DMC algorithm without our recent modifications. 
The DMC algorithm employs the importance-sampled exponential projector
\beq
\label{eq:Green}
G(\Rvecp,\Rvec,\tau) &=& {\PsiTRp \over \PsiTR} \brakett{\Rvecp}{e^{\tau(\ET-\hat{H})}}{\Rvec}
\eeq
at each Monte Carlo step to project onto the fixed-node ground state.
For a system of $N$ electrons, we use lowercase letters to denote $3$-dimensional vectors which correspond to a single electron, and
uppercase letters to denote $3N$-dimensional vectors which correspond to the entire electron configuration. 
For a given Monte Carlo step, we let $\rveci$ and $\rvecip$ be the initial and final positions of the $\ith$ electron and let $\Rvec$ and $\Rvecp$ be the initial and final electron configurations of the walker, respectively.

A Monte Carlo step consists of a stochastic step followed by a reweighting step.
During the stochastic step, electrons are moved one-by-one in a loop.
Suppose we have already moved the first $i-1$ electrons and are about to move the $\ith$ electron.
We first \textit{propose} a one-electron position $\rvecipp$ with probability
\beq
\Ppropf = \left( \frac{1}{2\pi\tau} \right)^{3N/2} \exp{ \left( {-\left( \rvecipp - \rveci - \vbarveciRi \tau \right)^2 \over 2\tau} \right) },
\label{eq:Gstoch}
\eeq
where
$$
\Rveci = \left( {\bf r}'_1, ..., {\bf r}'_{i-1}, {\bf r}_i, ..., {\bf r}_N \right)
$$
and
$$
\Rvecipp = \left( {\bf r}'_1, ..., {\bf r}'_{i-1}, {\bf r}''_i, ..., {\bf r}_N \right)
$$
are the initial and proposed final electron configurations for this iteration of the loop, respectively.
The average one-electron drift velocity $\vbarveciR$, introduced by Umrigar et al. \cite{UmrNigRun-JCP-93}, is
\beq
\vbarveciR = \left( \frac{-1 + \sqrt{1 + 2a \left\lvert \vveciRi \right\rvert^2 \tau}}{a \left\lvert \vveciRi \right\rvert^2 \tau} \right) \vveciRi,
\label{eq:vav}
\eeq
where the instantaneous one-electron drift velocity is
\beq
\vveciR = \frac{1}{\PsiT\left( \Rvec \right)} {\bf \nabla}_i \PsiT\left( \Rvec \right),
\eeq
and ${\bf \nabla}_i$ is the $3$-dimensional gradient of the $\ith$ electron.
Eq.~\ref{eq:vav} is derived using a simple local ansatz~\cite{AndUmr-JCP-21} for the trial wave function to determine the average drift velocity over the time step.
If the second derivative of $\PsiT$ along the direction of $\vveciR$ is calculated, the value of $a$ can be used to ensure that the second
derivative of the ansatz matches this value, otherwise $a$ is set to an adhoc constant, typically 0.5
for systems employing pseudopotentials.
The one-electron move from $\Rveci$ to $\Rvecipp$ is then accepted with probability~\cite{ReyCepAldLes-JCP-82}
\beq
\Pacc = \min\left\{ 1, \frac{\left\lvert \PsiT\left(\Rvecipp \right) \right\rvert^2 \Ppropb}{\left\lvert \PsiT\left( \Rveci \right) \right\rvert^2 \Ppropf}\right\}
\label{eq:P_acc}
\eeq
in a Metropolis-Hastings accept-reject step.
If the proposed position is accepted then $\rvecip = \rvecipp$. Otherwise, the electron remains at its original position and $\rvecip = \rveci$.
The Metropolis-Hastings accept-reject step ensures that expectation values of all operators are exact in the
$\PsiT \to \Psiz$ limit, where $\Psiz$ is the exact wave function.
In this limit the reweighting factor (discussed next) is unity and the DMC algorithm reduces to a
variational Monte Carlo algorithm with a particular choice of the Metropolis-Hastings proposal probability.
Once this procedure has been repeated for each of the $N$ electrons, the final electron configuration $\Rvecp$ is completely determined and the stochastic step is complete.

The reweighting step consists of multiplying the initial walker weight $w$ by a multiplicative reweighting factor to obtain the final weight $w' = w~\Delta w$, where
\beq
\Delta w = \exp \left( {\SRp + \SR \over 2 } \taueff \right).
\label{eq:dw}
\eeq
The naive expression for $\Delta w$ is obtained by setting $\taueff=\tau$ and
\beq
\SR = \ET - \ELR.
\label{eq:S_naive}
\eeq
This expression for
$\SR$ is accurate provided that $\ELR$ is approximately a constant in a $\sqrt{\tau}$ neighborhood of $\Rvec$.
For realistic systems, $\ELR$ is far from constant and in fact diverges to $\pm \infty$ at the nodal surface of the trial wave function.
As a walker moves its local energy decorrelates from the starting value, 
so a better choice is~\cite{UmrNigRun-JCP-93}
\beq
\SR = \ET-\Eest + \left(\Eest - \ELR \right) \fRt
\label{eq:S}
\eeq
where $\fRt$ is a suppression factor less than or equal to one which prevents the divergence in the local energy from causing a divergence in the walker weights.
One popular choice~\cite{UmrNigRun-JCP-93}, hitherto referred to as UNR93, is
\beq
\fUNRR = {\VbarR \over V(\Rvec)} = {\sqrt{\sumi \lvert \vbarveciR \rvert^2 } \over \sqrt{\sumi \lvert \vveciR \rvert^2} },
\label{eq:fUNR}
\eeq
and is derived using the same simple ansatz for the trial wave function as was used to obtain Eq.~\ref{eq:vav},
but to determine the average energy over a time-step, neglecting the motion due to diffusion.
However, for small time steps, the motion due to diffusion is larger than that from drift.
In the following section, we will mention several requirements that a good suppression factor $\fRt$ must satisfy, and introduce a specific suppression factor which meets these requirements and reduces the time-step error in the total energy.
We mention in passing that in the limit that $a \to 0$ (in Eq.~\ref{eq:vav}) the reweighting factor reduces to the naive reweighting factor,
whereas in the $a \to \infty$ limit the variational energy is recovered for finite $\tau$.

The effective time step $\taueff$ used in Eq.~\ref{eq:dw} is smaller than $\tau$ due to the fact that the Metropolis-Hastings accept-reject step reduces the expected distance diffused by each electron.
Without the accept-reject step, $\tau$ is equal to the expected squared distance diffused by each electron along each dimension when using Bohr units for distance and inverse Hartree units for imaginary time.
With the accept-reject step, this distance is reduced by an amount approximately equal to the rejection probability, and so a reasonable choice for $\taueff$ is
\beq
\taueff = \tau \frac{\sumi p_i (\delta r_i)^2 }{\sumi (\delta r_i)^2},
\label{eq:taueff}
\eeq
where $p_i = \Pacc$ is the acceptance probability of the $\ith$ one-electron move, and $(\delta r_i)^2 = \di^2$ is the squared distance diffused by the $\ith$ electron.

To avoid confusion we point out that the \emph{naive algorithm} we refer to in this paper employs the naive
reweighting of Eq.~\ref{eq:S_naive}, i.e. $f(\Rvec,\tau)=1$, but has two very important improvements compared to a truly
naive algorithm, namely, it utilizes the Metropolis-Hastings accept-reject step and $\taueff$ introduced in
Ref.~\onlinecite{ReyCepAldLes-JCP-82} and the average velocity introduced in Ref.~\onlinecite{UmrNigRun-JCP-93}.
In fact a truly naive algorithm that did not employ these would have infinite variance and would not be usable at all
for any system with nodes in the trial wave function.

\section{Reweight factor with small time-step errors}
\label{sec:rewt_factor}

The suppression factor, $\fRt$ in Eq.~\ref{eq:S}, expresses the fact that the expectation value of a walker's local energy evolves from $\ELR$ to $\EDMC$ as $\tau \to \infty$.
This is of particular importance when $\Rvec$ is near a node because the local energies of the approximate trial wave functions used in practice diverge to $+\infty$ on one side of the nodal surface and to $-\infty$ on the other side.
This is because as the distance $d(\Rvec)$ between the walker and the nodal surface goes to zero, $\PsiT(\Rvec)$ goes exactly to zero while $\nabla^2 \PsiT(\Rvec)$ becomes small but, for approximate trial wave functions, is not exactly zero.
This leads to a $1 \over d(\Rvec)$ divergence in the local kinetic energy.
The drift velocity also diverges for a similar reason and pushes the walker away from the nodal surface so that the local energy quickly relaxes
to a value closer to $\EDMC$.
To achieve small time-step errors, $\fRt$ must be chosen so as to accurately reproduce these and other behaviors even for large time steps.
We will propose an improved $\fRt$ based on the following five reasonable requirements.

\begin{enumerate}
\item 
We first note that as $\tau \to 0$ the reweighting factor $\SR$ must reduce to the naive expression, Eq.~\ref{eq:S_naive}, for which $\fRt=1$.
An ${\cal O}(\tau^n)$ modification to $G$ makes an ${\cal O}(\tau^{n-1})$ modification to the total energy because
${\cal O}(1/\tau)$ Monte Carlo steps are needed to project to a given physical time.
Hence, an ${\cal O}(\tau)$ deviation of $\fRt$ from one gives an ${\cal O}(\tau)$ increase in the total energy so that the slope of the energy versus time-step curve will increase at $\tau=0$.
Note that because $\fUNRR$ deviates from 1 linearly in $\tau$, it increases the slope at $\tau=0$ relative to the naive algorithm and its energy versus time-step curve
usually has a positive slope at $\tau=0$ and a hump at small $\tau$.
The height and extent of the hump get smaller as $a$ in Eq.~\ref{eq:vav} goes to zero.  For pseudopotential
systems we typically use $a=0.5$ in which case the hump is very noticeable, whereas for all-electron systems we use a
more complicated formula for $a$ described in Ref.~\onlinecite{UmrNigRun-JCP-93} and there is typically no noticeable hump.
On the other hand, for all systems tested, the naive reweighting in Eq.~\ref{eq:S_naive} (with the Metropolis-Hastings step,
the $\vbar$ of Eq.~\ref{eq:vav} and $\taueff$)
gives a slope at $\tau=0$ that appears to be
fairly small and negative (though it requires very long runs at very small values of $\tau$ to pin it down).
Since it is already small, we choose to not alter the slope at $\tau=0$.
\vspace{-3mm}
\begin{center}
{\bf Requirement 1:} For nonzero ${d(\Rvec)}$, ${\fRt \to 1 - {\cal O}(\tau^2)}$ as ${\tau \to 0}$.
\end{center}

\vskip 1mm
\item For finite $\tau$, as the distance $d(\Rvec)$ between $\Rvec$ and the nearest node goes to $0$, the reweighting factor should go to
a constant greater than 1 if the walker starts on the side of the node where $\ELR \to -\infty$ and to
a constant less than 1 if the walker starts on the side of the node where $\ELR \to +\infty$.
To accomplish this, $\fRt$ must go to zero linearly in $d(\Rvec)$ in order to cancel the $\pm \frac{1}{d(\Rvec)}$ divergence in $\ELR$.
\vspace{-3mm}
\begin{center}
{\bf Requirement 2:} For nonzero ${\tau, \; \fRt \propto d(\Rvec)}$ as ${d(\Rvec) \to 0}$.
\end{center}

\vskip 1mm
\item The local energy of a walker decorrelates from its initial value as the walker evolves and at large
$\tau$ its weighted average is the DMC energy.
Consequently, $\Delta w$ should go monotonically to a finite constant
(rather than 0 or infinity for the naive expression) as $\tau$ increases.
So, for large $\tau$ the deviation of the average energy during $\tau$ from the DMC energy must go down monotonically as $1/\taueff$.
In contrast $\fUNRR$ goes down only as $1/\sqrt{\tau}$, which is likely the reason that
$\fUNRR$ typically has a large negative time-step error at large $\tau$.
\vspace{-3mm}
\begin{center}
{\bf Requirement 3:} ${\fRt \sim 1 / \taueff}$, at large ${\tau}$.
\end{center}

\vskip 1mm
\item When the walker is close to a node, the drift pushes it away from the node
and $\ELR$ very quickly decorrelates from its initial near-divergent value as a function of $\tau$.
\vspace{-3mm}
\begin{center}
{\bf Requirement 4:} As $d(\Rvec) \to 0$, $\fRt$ should decay from 1 to 0 more quickly in $\tau$.
\end{center}

\vskip 1mm
\item It is desirable to have a size-consistent algorithm, i.e., the energy of a system consisting of
$M$ widely separated fragments should equal the sum of the fragment energies.
In the next section, we will discuss modifications which make the algorithm {\emph{precisely}} size-consistent and
are useful for reducing the time-step error in the binding energy of some weakly bound systems.
Here we present an {\emph{approximately}} size-consistent requirement which should be useful for reducing
the time-step error in the total energy of any system.

The DMC algorithm is exactly size-consistent for a system of $M$ widely separated fragments if the trial wave function of the composite system is a product of the trial wave functions of each fragment and the reweighting factor satisfies
\beq
\dw = \prodk \dwk,
\label{eq:rewt_prod}
\eeq
where $\dw$ is the total multiplicative reweighting factor of the entire system and $\dwk$ is the multiplicative reweighting factor obtained from the $k^{\rm th}$ fragment alone.
Note that for independent fragments we regard the full electron configuration as being the union of the electron configurations of each fragment alone so that $\Rvec = \left( \Rvec_1, ..., \Rvec_{M} \right)$.
Writing Eq.~\ref{eq:rewt_prod} in terms of the suppression factor yields
\beq
\left(\ET - \ELRvec \right) f(\Rvec,\taueff) \taueff = \sumk \left( \ETk - \ELkRvec \right) f_k(\Rvec_k,\taueffk) \taueffk,
\label{eq:f_additive}
\eeq
where $\ETk$, $\ELkRvec$, and $\taueffk$ are analogous to corresponding quantities for the composite
system, but for the $k^{\rm th}$ fragment.
Note that for independent fragments if we choose $\ET = \sumk \ETk$ then we will also have $\ET - \ELRvec = \sumk \left( \ETk - \ELkRvec \right)$.
Also note that if we set $\taueff=\taueffk=\tau$ then the naive reweighting factor, corresponding to $\fRt=f_k(\Rvec_k,\taueffk)=1$, is perfectly size-consistent and already satisfies Eq.~\ref{eq:f_additive} exactly.
However, this of course is not a viable option because of the large
negative time-step errors in the total energies and the occurrence of population explosions.

Although we cannot otherwise satisfy Eq.~\ref{eq:f_additive} in general, we can require that it be satisfied in the special case of $M$ {\emph{identical}} widely separated fragments with identical electron configurations.
Because the fragments are identical we have $\taueffk=\taueff$.
Because the electron configurations are also identical, can arrange that $f(\Rvec,\taueff)= f_k(\Rvec,\taueffk)$ as follows.
The distance to the nearest node can be estimated as $d(\Rvec)= {\lvert \PsiR \rvert \over \lvert {\bf \nabla} \PsiR \rvert } = {1 \over V(\Rvec)}$.
Hence one can use ${1 \over V(\Rvec)}$ instead of $d(\Rvec)$ in $\fRteff$.
Now, $V(\Rvec)$ for the composite system increases as the square root of the number of identical fragments.
Hence, in this very special case, Eq.~\ref{eq:f_additive} will be satisfied if $f$ is
written in terms of ${V(\Rvec) \over \sqrt{N}}$, where $N$ is the number of electrons in the system
(either composite or fragment).
More generally (i.e., for non-identical fragments or non-identical electron configurations) Eq.~\ref{eq:f_additive}
will be only approximately satisfied.
For non-identical electron configurations, one can numerically verify that Eq.~\ref{eq:f_additive}
can be satisfied by replacing $\sqrt{N}$ by a slightly smaller electron-configuration dependent power of $N$, with some assumptions, e.g.,
that $\tau$ is small.  However, in this paper we do not explore this avenue for further improvement.
\vspace{-3mm}
\begin{center}
{\bf Requirement 5:} ${\fRt}$ should depend on the combination ${V(\Rvec) \over \sqrt{N}}$.
\end{center}

\end{enumerate}

\def\fimpRt{f_{\rm imp}\left(\Rvec,\tau\right)}

Two expressions for $\fRt$ that satisfy the above 5 requirements are
\beq
\fimpRt &=& \left( 1 + \left({c ~ V(\Rvec) \taueff \over \sqrt{N}}\right)^2 \right)^{-1/2},
\label{eq:new_rewt1}
\eeq
\vskip -10mm
and
\vskip -15mm
\beq
\fimpRt &=& {\left({\sqrt{\pi} \over 2}\right) \erf\left({cV(\Rvec)\taueff \over \sqrt{N}}\right) \over \left({cV(\Rvec)\taueff \over \sqrt{N}}\right)}.
\label{eq:new_rewt2}
\eeq
We will employ Eq.~\ref{eq:new_rewt2} in all our calculations in the body of the paper, although we found that the difference between the energies from
Eq.~\ref{eq:new_rewt1} and \ref{eq:new_rewt2} is much smaller than the difference from other choices for $\fRt$.
In the supplementary material we show energy versus $\tau$ curves for three all-electron (no pseudopotential) systems,
Be, Ne and N$_2$, using Eq.~\ref{eq:new_rewt1} to emphasize that we have no reason to prefer either of these choices.

We note that both $V(\Rvec)$ and $|\Eest-\EL(\Rvec)|$ diverge as $1/d(\Rvec)$ at wave function nodes.
Consequently, other reasonable choices of $\fimpRt$ can be obtained by replacing $V(\Rvec)/\sqrt{N}$ by
either $|\Eest-\EL(\Rvec)|/\sqrt{N}$ or $|\Eest-\EL(\Rvec)|/\sigma_E$, where $\sigma_E$ is the root mean square fluctuation of
the local energy.  Either $\sqrt{N}$ or $\sigma_E$ are introduced here to again ensure approximate size-consistency.
Finally we note that an argument, similar to our requirement 5, has previously been used by Zen et al.~\cite{ZenSorGilMicAlf-PRB-16}
to propose the sharp cutoff factor $\fRt = \min\left(1,{{0.2 \sqrt{N/\tau}} \over |\Eest-\EL(\Rvec)|}\right)$.

In Eqs.~\ref{eq:new_rewt1} and \ref{eq:new_rewt2},
$c$ is a constant which determines the rate at which $S(\Rvec)$ decorrelates from its initial value $\ET-\ELR$, and is thus related to the autocorrelation time $T_{\rm corr}$ of the local energy of the system.
$S(\Rvec)\taueff$ tends to a constant as $\tau$ increases,
and the absolute value of the constant is both smaller and is approached more
rapidly the larger the value of $c$.  Hence we expect that optimal values of $c$ will be large for
systems with small autocorrelation times, and small for systems with large autocorrelation times.
We empirically found a near optimal $c$ for one particular system, which we chose to be a carbon atom
using the Burkatski-Filippi-Dolg (BFD) pseudopotential~\cite{BurFilDol-JCP-08}, a single
configuration state function (1-CSF) wave function, and $\tau=0.01$ Ha$^{-1}$.
This value is $c^{\rm Carbon-1CSF}=3.5$.
Then, for other systems we can predict a reasonable choice for $c$ by computing its autocorrelation time, $\Tcorr$,
in units of the number of Monte Carlo steps for a run with $\tau=0.01$ Ha$^{-1}$, using the formula
\beq
c = c^{\rm Carbon-1CSF} \sqrt{T^{\rm Carbon-1CSF}_{\rm corr} - 1 \over T_{\rm corr} - 1} \;=\; {k \over \sqrt{T_{\rm corr} - 1}},
\label{eq:c_x}
\eeq
where $k=15.51$ for pseudopotential systems.
The ``-1" in Eq.~\ref{eq:c_x} reflects the fact that we use a definition of the autocorrelation time for
which $\Tcorr=1$ (rather than 0) for uncorrelated energies
($\Tcorr = 2\tcorr+1$, where $\tcorr$ is the usual definition of the integrated autocorrelation time).
The autocorrelation time for $\tau=0.01$ Ha$^{-1}$ for each system is obtained from a short DMC run, using a single walker,
with the reweighting turned off, i.e., it is a VMC run that uses the same proposal probability density
as DMC.  The computational cost for this extra step is negligible compared to the cost of the
actual DMC calculation.

The values of $T_{\rm corr}$ and $c$ for the various systems computed in this paper are shown
in Table~\ref{tab:c_x}.
Autocorrelation times depend not only on the system but also on the quality of the wave function.
Multideterminant wave functions have somewhat smaller autocorrelation times than single-determinant wave functions,
because not only are the magnitudes of the local energy fluctuations smaller but they also fluctuate more rapidly in space.
For example, the 29-CSF wave function for the C atom in Table~\ref{tab:c_x} and Fig.~\ref{fig:total_energy_timestep}
has a smaller autocorrelation time than the 1-CSF wave function.
Similarly, atoms with more core electrons tend to have more rapid fluctuations of the local energy and therefore smaller values of $\Tcorr$
as can be seen for Cr in Table~\ref{tab:c_x}.

Of course the values of $c$ given by Eq.~\ref{eq:c_x} are not optimal, but they are reasonably close,
e.g., Eq.~\ref{eq:c_x} gives $c=8.1$ for Cr, but empirically a better value is $c=8.5$,
and, for Si$_{15}$, Eq.~\ref{eq:c_x} gives $c=2.5$ but $c=2.9$ is better.

\begin{table}[htbp]
\caption{Autocorrelation times, in units of Monte Carlo steps, for $\tau=0.01$ Ha$^{-1}$ for various pseudopotential systems and $c$
according to Eq.~\ref{eq:c_x}. Unless otherwise stated, 1-CSF wave functions are used. When using the fragments approach presented in Sec.~\ref{sec:fragment}, a separate value of c is calculated for each fragment.}
\begin{tabular}{|l|d|d|d|d|d|d|}
\hline
System (pseudopotential type)          &\multicolumn{1}{c|}{$\Tcorr$} &\multicolumn{1}{c|}{$\Tcorr$} &\multicolumn{1}{c|}{$\Tcorr$} & \multicolumn{1}{c|}{$c$} & \multicolumn{1}{c|}{$c$} & \multicolumn{1}{c|}{$c$} \\
                            &\multicolumn{1}{c|}{system}   &\multicolumn{1}{c|}{frag 1  } &\multicolumn{1}{c|}{frag 2  } & \multicolumn{1}{c|}{system} & \multicolumn{1}{c|}{frag 1  } & \multicolumn{1}{c|}{frag 2  } \\
\hline
C  1-CSF (BFD)                         & 20.63  &       &       & 3.50 &      &       \\
C 29-CSF (BFD)                         & 14.96  &       &       & 4.15 &      &       \\
Cr (BFD)                               &  4.68  &       &       & 8.08 &      &       \\
H$_2$O (BFD)                           & 12.38  &       &       & 4.60 &      &       \\
Butadiene (BFD)                        & 22.26  &       &       & 3.36 &      &       \\
Si$_{15}$ (BFD)                        & 40.05  &       &       & 2.48 &      &       \\
H$_2$O (ccECP)                         & 11.47  &       &       & 4.79 &      &       \\
CH$_4$ (ccECP)                         & 21.52  &       &       & 3.42 &      &       \\ 
H$_2$O-CH$_4$ (ccECP)                  & 15.35  & 12.41 & 24.66 & 4.09 & 4.59 & 3.19  \\
H$_2$O-\hskip .8mm-\hskip .8mm-CH$_4$ (ccECP)                & 14.19  & 11.47 & 21.52 & 4.27 & 4.79 & 3.42  \\
\hline                                               
\end{tabular}
\label{tab:c_x}
\end{table}

\newpage
\section{Modified Reweight Factor for Weakly Bound Systems}
\label{sec:fragment}

In this section, we present a modification to the reweighting factor which reduces the time-step error in the interaction energies of weakly interacting systems.
We do this by devising a reweighting factor which is perfectly size-consistent and thus has zero time-step error in the binding energies of noninteracting
systems.
We will refer to the weakly interacting systems as ``fragments" of the composite system.

The fragments are defined at the beginning of the computation by specifying which nuclei belong
to each fragment.  For noninteracting systems each electron will stay on the same fragment for the entire
run, but for weakly interacting systems electrons can migrate from one fragment to another.
For neutral systems, each nucleus, $\alpha$, has $Z_\alpha$ electrons assigned to it.
For ionic systems, it is necessary to decide how many electrons are assigned to each atom or each fragment.
At the beginning of the reweighting step of each Monte Carlo step, each electron is first assigned to a particular nucleus and then assigned to that nucleus' fragment.
We assign electron $i$ to nucleus $\alpha(i)$ by maximizing
\beq
C_{\rm tot} = \sum^{N}_{i=1} {Z_{\alpha(i)} \over r_{i \alpha(i)}}
\eeq
over all possible assignments.  Although the number of possible assignments is combinatorially large,
it can be achieved in $\mathcal{O}(N^3)$ time, where $N$ is the number of electrons, using the Hungarian algorithm~\cite{HungarianAlg}.
Since the time complexity of a standard Monte Carlo step is already $\mathcal{O}(N^3)$ time and we use the
Hungarian algorithm only once per Monte Carlo step, the overall time complexity of the DMC algorithm is left
unchanged and in fact the increase in computer time is very minor.

For a system with $N$ electrons and $N_{\rm nucl}$ nuclei, the potential energy of a given walker is
\beq
\sum_{i=1}^{N+N_{\rm nucl}} \sum_{j \neq i} {\cal V}(\rij),
\eeq
where the pairwise potential, ${\cal V}$, depends on the nature of the interacting particles and is either a Coulomb or a
pseudopotential interaction.
For noninteracting fragments this can be written as a sum of intra-fragment terms.
For interacting fragments, the inter-fragment terms, regardless of the nature of the two interacting
particles, are split equally among the two fragments to which the two particles belong when calculating the local energy $\ELkRvec$ of those fragments.

The DMC energy of the composite system will equal the sum of the DMC energies of noninteracting fragments if
\beq
\Delta w(\Rvec,\tau) = \prodk \Delta w_k(\Rvec_k,\tau),
\label{eq:dwprod}
\eeq
where $\Delta w_k$ is the reweighting factor that fragment $k$ would have if it were computed in isolation.
Hence, in addition to the Hamiltonian, the variables that appear in the reweighting factor in
Eqs.~\ref{eq:dw}, \ref{eq:S}, \ref{eq:fUNR}, \ref{eq:taueff}, \ref{eq:new_rewt2}
namely, $\ET,\,\Eest,\,\EL,\,V,\,\Vbar,\,\taueff,$ and $c$, acquire a fragment index and need to be evaluated for each fragment.
This requires some book keeping, but it adds negligibly to the cost, ensures that the interaction
energy of noninteracting systems is precisely zero, and reduces the time-step error for the interaction
energy of some weakly interacting systems, as we show in the results section.

\section{Computational Details}
\label{sec:details}

All the calculations employ Jastrow-Slater wave functions.  Since the goal is to study the improvement in
the time-step error, we intentionally do not employ the best trial wave functions we can generate --
we use small basis sets and
in all except one system we employ just single determinant wave functions.
The orbitals are optimized for the systems shown in Fig.~\ref{fig:total_energy_timestep} but not
for the systems shown in Figs.~\ref{fig:total_energy_timestep_H2O_CH4} and \ref{fig:bind_energy_timestep}.
The systems in Figs.~\ref{fig:total_energy_timestep_H2O_CH4} and \ref{fig:bind_energy_timestep} are H$_2$O, CH$_4$,
the weakly interacting H$_2$O-CH$_4$ dimer at equilibrium distance, and widely separated H$_2$O and CH$_4$ 
(denoted by H$_2$O-\hskip 1mm -\hskip 1mm -CH$_4$).  The geometries of all the systems are provided in the ancillary files~\cite{geo}.

The 3-d version of the 5$^{th}$-order Jastrow factor of the form used in Ref.~\onlinecite{GucJeoUmrJai-PRB-05}
was employed.
The Jastrow factor contains electron-nuclear (e-n), electron-electron (e-e) and electron-electron-nuclear (e-e-n) terms.
The e-n and e-e-n terms are atomic-species dependent.
In order to get good size-consistency for H$_2$O-\hskip 1mm -\hskip 1mm -CH$_4$ and a small time-step error for the binding energy
of H$_2$O-CH$_4$, we found it necessary to use different Jastrow parameters for the hydrogen atoms in water and in methane.
Using the same e-n and e-e-n Jastrow parameters for the hydrogen atoms in H$_2$O and CH$_4$ resulted
in a size-consistency error of $2.5 \pm 0.2$ mHa for H$_2$O-\hskip 1mm -\hskip 1mm -CH$_4$ in variational Monte Carlo (VMC).
Upon allowing these to be different, the size-consistency error is reduced to $0.2 \pm 0.2$ mHa, i.e.,
there is no statistically significant size-consistency error.
Even with this additional variational freedom, H$_2$O-\hskip 1mm -\hskip 1mm -CH$_4$ has slightly less
variational freedom than H$_2$O and CH$_4$ calculated separately because there is a single e-e term
in H$_2$O-\hskip 1mm -\hskip 1mm -CH$_4$, whereas when H$_2$O and CH$_4$ are optimized separately they
each have their own e-e terms.
Hence, when we show the DMC size-consistency error in Section~\ref{sec:results} we present separate curves, one with
the independent optimization of the Jastrow factors for each of H$_2$O, CH$_4$, and H$_2$O-\hskip 1mm -\hskip 1mm -CH$_4$
and another with optimization of the Jastrow factor of just H$_2$O-\hskip 1mm -\hskip 1mm -CH$_4$. 
In the latter case,
the appropriate pieces of the H$_2$O-\hskip 1mm -\hskip 1mm -CH$_4$ Jastrow factor are used for the separate
H$_2$O and CH$_4$ systems.  Consequently, the wave function for
H$_2$O-\hskip 1mm -\hskip 1mm -CH$_4$ is a product of the H$_2$O and CH$_4$ wave functions,
and our fragment algorithm must be exactly size-consistent.

For the systems in Fig.~\ref{fig:total_energy_timestep}, the Burkatski-Filippi-Dolg pseudopotentials~\cite{BurFilDol-JCP-08}
(with a corrected potential for H provided by the authors) with a double-zeta basis
were employed, whereas for the systems in Figs.~\ref{fig:total_energy_timestep_H2O_CH4} and \ref{fig:bind_energy_timestep} the ccECP pseudopotentials of
Refs.~\onlinecite{BenMelAnnWanShuMit-JCP-17,Mit-PseudopotentialLibrary} with a triple-zeta basis were employed.
These nonlocal pseudopotentials were treated using our modified T-moves method~\cite{AndUmr-JCP-21}.
This approximation has been shown to give smaller time-step errors than
the original T-moves methods of Casula and coworkers~\cite{Cas-PRB-06,CasMorSorFil-JCP-10},
particularly for expectation values of observables that do not commute with the Hamiltonian.

\section{Results}
\label{sec:results}

\begin{figure}[htbp]
\centering
\subfigure[]{{\includegraphics[width=3.5in,height=2.8in,clip]{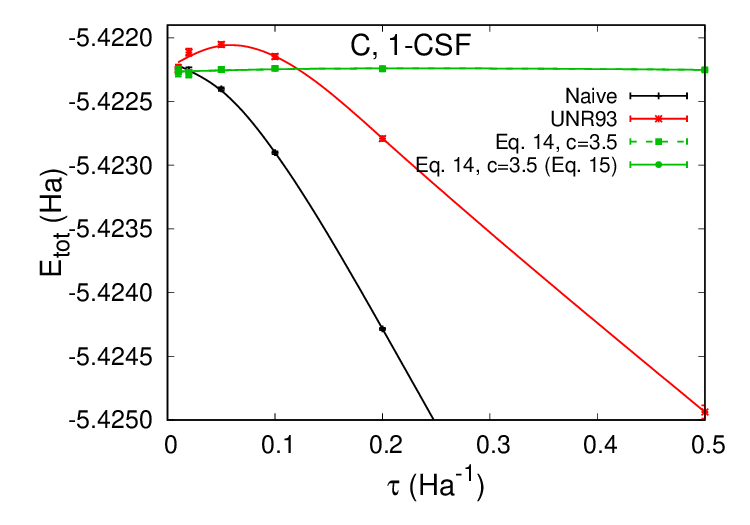}}}\quad
\subfigure[]{{\includegraphics[width=3.5in,height=2.8in,clip]{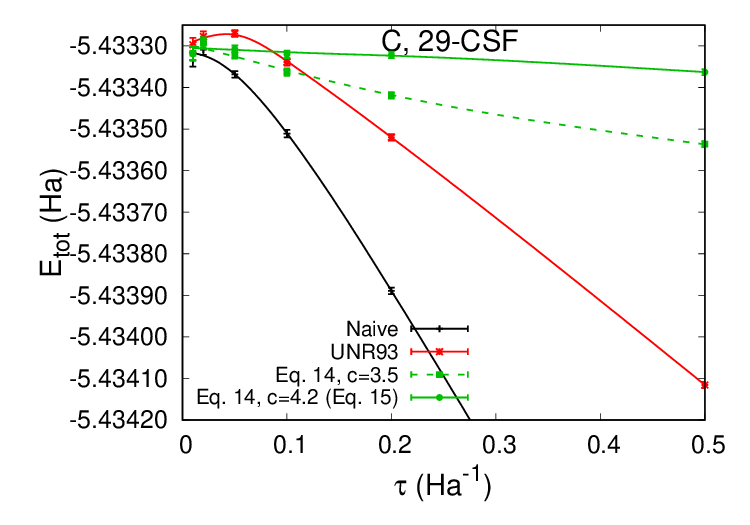}}}\quad\\
\subfigure[]{{\includegraphics[width=3.5in,height=2.8in,clip]{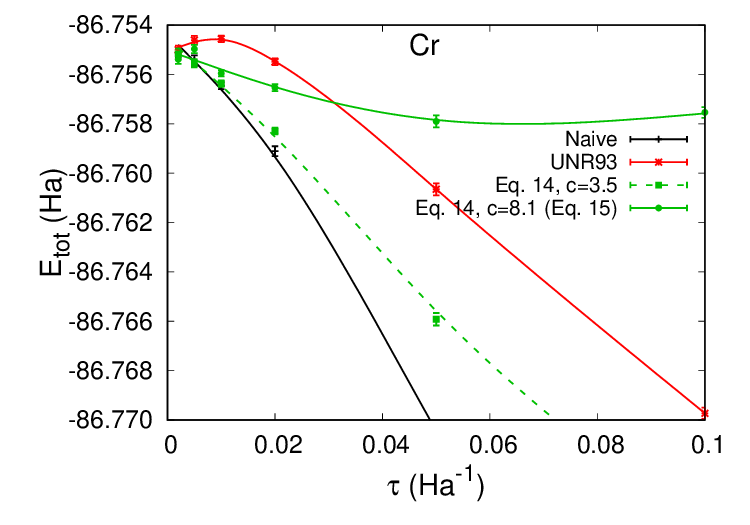}}}\quad
\subfigure[]{{\includegraphics[width=3.5in,height=2.8in,clip]{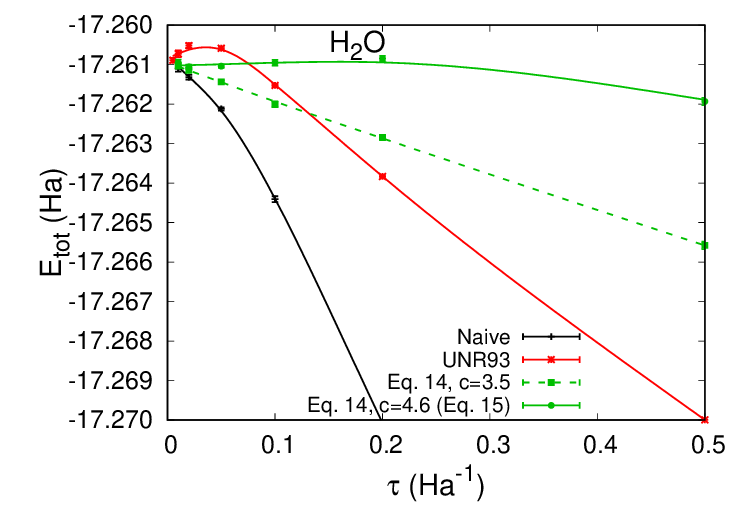}}}\quad
\subfigure[]{{\includegraphics[width=3.5in,height=2.8in,clip]{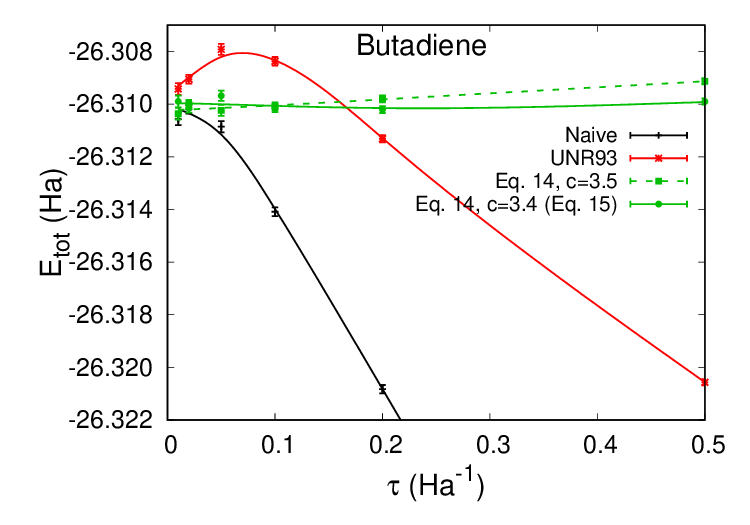}}}\quad
\subfigure[]{{\includegraphics[width=3.5in,height=2.8in,clip]{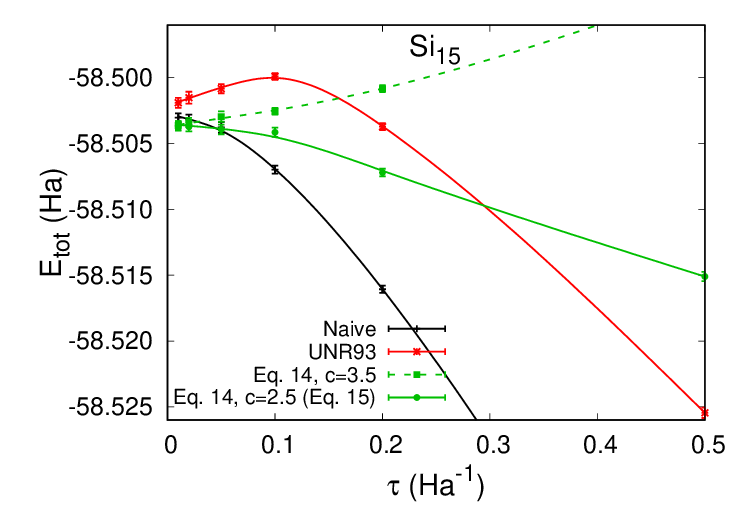}}}\quad
\caption{Comparison of time-step errors of the total energy for the naive, UNR93 and new reweighting factors.
The dashed and solid green curves employ the new reweighting with $c=3.5$ and $c$ from Eq.~\ref{eq:c_x} respectively.
}
\label{fig:total_energy_timestep}
\end{figure}

In Fig.~\ref{fig:total_energy_timestep} we compare the time-step errors of the total energies of several systems for 4 different reweighting factors:
the naive expression corresponding to $f=1$ in Eq.~\ref{eq:S}, the UNR93 expression corresponding to the $f$ of Eq.~\ref{eq:fUNR},
and the improved expression corresponding to the $f$ of Eq.~\ref{eq:new_rewt2} with $c=3.5$ and with $c$ from Eq.~\ref{eq:c_x}.
The systems were selected, in part, to have a wide range of autocorrelation times, $\Tcorr$ (short for Cr, long for Si$_{15}$), to test how well
the prescription of Eq.~\ref{eq:c_x} for selecting the value of $c$ works.
Note that the time steps in Fig.~\ref{fig:total_energy_timestep} are an order of magnitude larger than those commonly used.
The naive expression has a strongly negative time-step error and it is known to be prone to population explosions
at very large time steps.
The UNR93 expression gives a smaller time-step error at large values of the time step and does not give population
explosions but it has a positive hump at small time steps that makes the curve difficult to extrapolate to
$\tau=0$.
The improved expression of Eq.~\ref{eq:new_rewt2} with $c$ from Eq.~\ref{eq:c_x} gives the smallest time-step error.
In the case of the 1-CSF carbon atom, this is in part because we chose the value of $c$ in Eq.~\ref{eq:new_rewt2}
to get a flat curve, but even for this case it should be noted that most other functional forms for
$f$ with one adjustable parameter will not give such a flat curve for any value of the parameter.
The curves using $c=3.5$ have in most cases a larger time step dependence than those with $c$ from Eq.~\ref{eq:c_x},
but even they are easier to extrapolate than the UNR93 curves.

Note that the plot for the 29-CSF carbon atom employs a much finer energy scale than the plot for the 1-CSF carbon atom,
because, as expected, improved wave functions give not only lower DMC energies but also smaller time-step errors.
Similar improvements in the time-step error could be obtained for the other systems, but we chose not
to do that because
the point of this paper is a \emph{relative} comparison of the time-step errors for different reweighting formulas,
regardless of the quality of the trial wave function.
Finally, we provide plots of the kinetic energy versus the time-step in the supplementary material and note that the improvement is not as consistent or as large as for the total energy.

\begin{figure}[H]
\centering
\subfigure[]{{\includegraphics[width=3.5in,height=2.8in,clip]{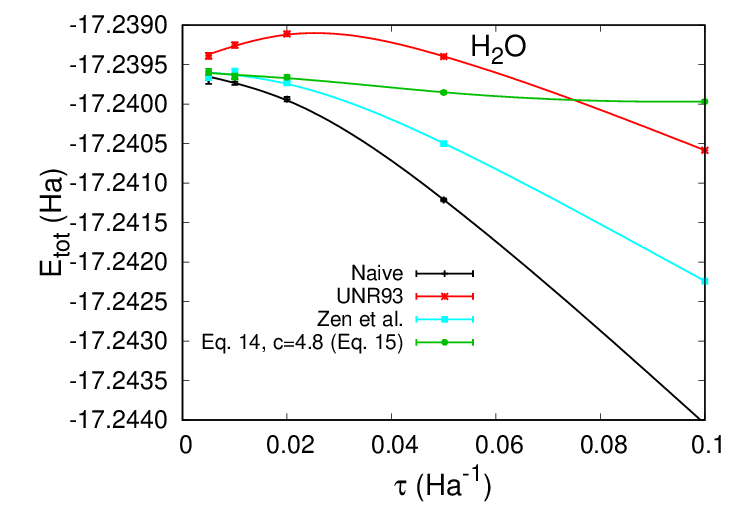}}}\quad
\subfigure[]{{\includegraphics[width=3.5in,height=2.8in,clip]{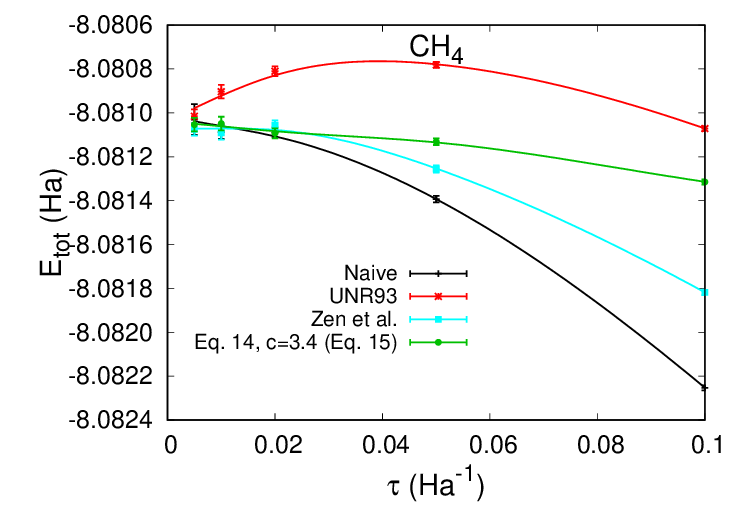}}}\quad\\
\subfigure[]{{\includegraphics[width=3.5in,height=2.8in,clip]{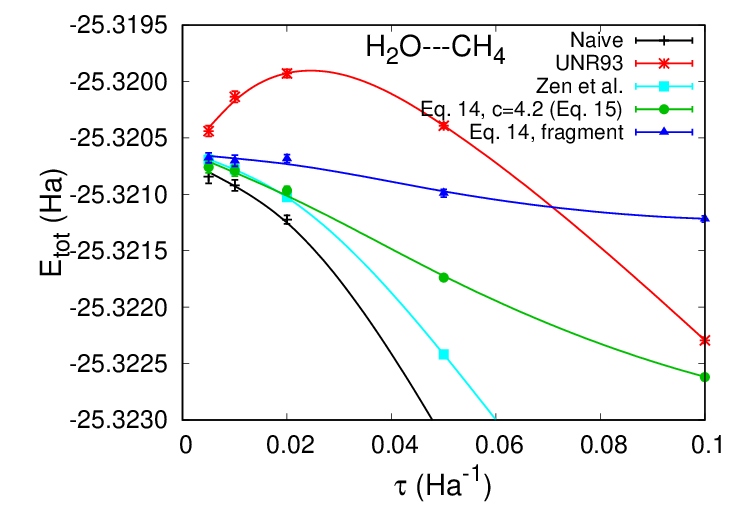}}}\quad
\subfigure[]{{\includegraphics[width=3.5in,height=2.8in,clip]{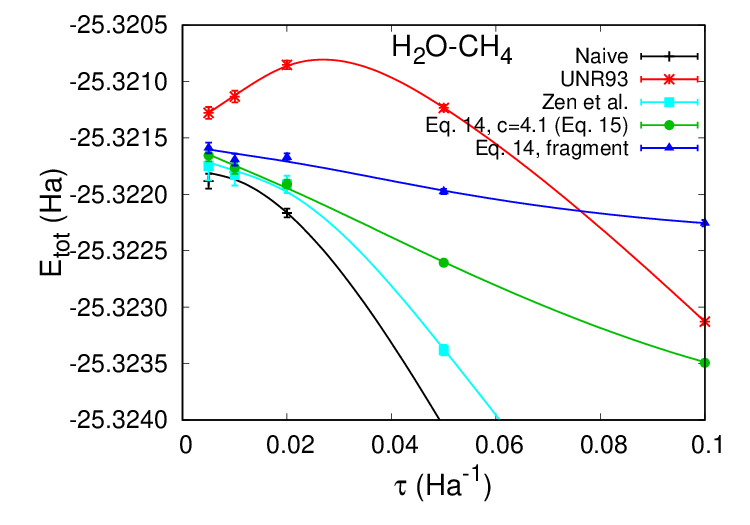}}}\quad
\caption{Comparison of time-step errors of the total energy for the naive, UNR93, Zen et al.~\cite{ZenSorGilMicAlf-PRB-16} and new reweighting factors.
For the composite systems, the errors of the new reweighting factor are shown both with and without the fragments approach.
}
\label{fig:total_energy_timestep_H2O_CH4}
\end{figure}

In Fig.~\ref{fig:total_energy_timestep_H2O_CH4} we show the time-step error for the total energies of water (H$_2$O), methane (CH$_4$),
and water-methane at both
wide separation (denoted H$_2$O-\hskip 1mm -\hskip 1mm -CH$_4$) and equilibrium separation (denoted H$_2$O-CH$_4$).
Similar improvements to the time-step error are seen as for the systems in Fig.~\ref{fig:total_energy_timestep}.

\begin{figure}[H]
\centering
\subfigure[]{{\includegraphics[width=3.5in,height=2.8in,clip]{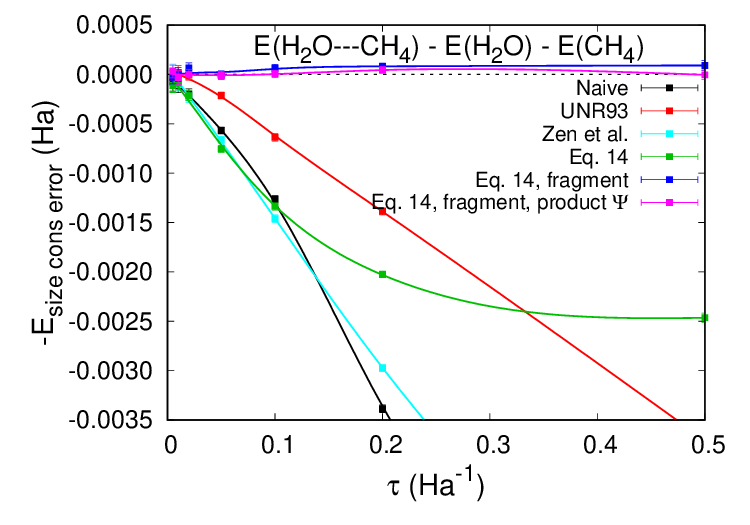}}}\quad
\subfigure[]{{\includegraphics[width=3.5in,height=2.8in,clip]{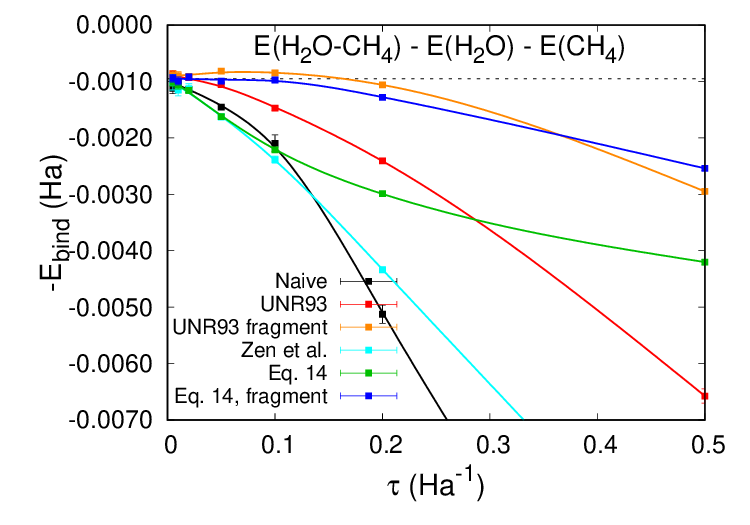}}}\quad\\
\caption{Comparison of size-consistency error and the binding energy errors as a function of the time step for
the naive, UNR93, and new reweighting factors.  The errors, using the new reweighting factor, are shown both with and without the fragments approach.
For the size-consistency error, the blue curve uses Jastrow factors that were independently optimized for the
three systems, while the magenta curve uses the appropriate pieces of the Jastrow factor optimized for the
composite system in the calculations for both H$_2$O and CH$_4$. Because of this, the wave function for the composite system is strictly a product
of the H$_2$O and CH$_4$ wave functions.
The former shows a very tiny size-consistency error at large $\tau$, whereas the latter has no
statistically significant size-consistency error.
Note that in the latter curve the largest deviation from zero is for $\tau=0.2 ~ {\rm Ha}^{-1}$, 
and equals $(4.5 \pm 3.2) \times 10^{-5} ~ {\rm Ha}$, so the size-consistency error is within two standard deviations of zero for all points.
}
\label{fig:bind_energy_timestep}
\end{figure}

In Fig.~\ref{fig:bind_energy_timestep} we subtract the energies of H$_2$O and CH$_4$ from the energies of the
combined systems to study the size-consistency error (Fig.~\ref{fig:bind_energy_timestep}$a$)
and the time-step error of the binding energy (Fig.~\ref{fig:bind_energy_timestep}{\it b}).
For this system, although the time-step error of the total energy from the new reweighting scheme is much smaller than
that from UNR93, the new reweighting gives a larger size-consistency and binding energy error.
Once we add in the fragment approach, the size-consistency error entirely disappears (magenta curve) provided that
the wave function of the composite system is a product of the wave functions of the individual systems.
If instead we independently optimize all the parameters in the composite system wave function
then there is a very tiny residual size-consistency error (blue curve).
The size-consistency plot is of course of no practical utility -- it is done simply to test the algorithm.
The usefulness of the algorithm is shown in Fig.~\ref{fig:bind_energy_timestep}{\it b} -- the time-step
error of the binding energy is reduced by using the fragment approach and shows no statistically
significant error up to $\tau=0.1$ (Ha$^{-1}$).
At equilibrium separation of course a product wave function is not a good approximation to the true wave function,
so the binding energy plot uses only an independently optimized wave function.

\section{Conclusions and Prospects for Further Improvements}
\label{sec:conclusions}

We have presented two modifications to the reweighting factor in DMC.  The first is generally applicable
and reduces the time-step error in the total energy for all systems tested.  It required only a trivial
modification of existing codes, but has one adjustable
parameter, which is determined by short auxiliary calculations involving just a single walker.
The second modification is of more limited utility -- it reduces the time-step error in the binding
energy for some weakly bound systems, and it requires a somewhat larger change to existing codes.

Further improvement may be possible.
From Figs.~\ref{fig:total_energy_timestep_H2O_CH4} and \ref{fig:bind_energy_timestep} it is apparent that better
size-consistency and a lower time-step error in the binding energy could be obtained, when the fragments approach is not used,
if requirement 5 was modified so that the $\sqrt{N}$ in Eqs.~\ref{eq:new_rewt1} and \ref{eq:new_rewt2}
was replaced by a somewhat smaller power of $N$, in agreement with the discussion provided there.
We have observed the same trend for some other systems as well.
This change would have the effect of increasing the slope of the energy versus $\tau$ curve for systems with
a large number of electrons compared to systems with few electrons.
We note that in Fig.~\ref{fig:total_energy_timestep} the system with the largest number of electrons, namely Si$_{15}$, is also the system
with the most negative slope for the new reweighting, so this modification would result in flatter total energy
curves also.

\section{Supplementary Material}
We provide plots of the total energy versus the time-step for three all-electron systems, Be (2 CSF), Ne (442 CSF), and N$_2$ (1880 CSF), 
in Fig.~S1 of the supplementary material (SM).
In the paper we use the reweighting Eq.~\ref{eq:new_rewt2} but in the SM we use Eq.~\ref{eq:new_rewt1} because there is no
theoretical reason to prefer either choice.
We observe an even larger reduction in the time-step error for these all-electron systems than we did for many of the
pseudopotential systems.
In Fig.~S2, we demonstrate that Eqs.~\ref{eq:new_rewt1} and \ref{eq:new_rewt2} do in fact give very similar curves
and that Eq.~\ref{eq:new_rewt1} in fact gives a slightly flatter curve than Eq.~\ref{eq:new_rewt2}.
We also provide plots of the kinetic energy versus the time-step, but note that the improvement is not as consistent or as large as for the total energy.

\section{Data Availability}
The data that support the findings of this study are available within the article, the supplementary material of
the arXiv version of this paper and from the corresponding authors.

\begin{acknowledgments}
This work was initiated under AFOSR (Grant No. FA9550-18-1-0095) and completed under the Exascale Computing Project (17-SC-20-SC),
a collaborative effort of the U.S. Department of Energy Office of Science and the National Nuclear Security Administration."
Some of the computations were performed at the Bridges cluster at the Pittsburgh Supercomputing Center supported
by the NSF (Grant No. ACI-1445606).
We thank Andrea Zen for useful comments on the manuscript.
\end{acknowledgments}

\clearpage

\bibliographystyle{apsrev4-1}
\bibliography{umrigar,sorella,biblio,mitas,needs,qmc,fragments}

\end{document}